\documentclass[12pt,preprint]{aastex}

\newcommand{\be}{\begin{enumerate}}
\newcommand{\ee}{\end{enumerate}}
\newcommand{\dg}{$^{\circ}$}
\newcommand{\um}{$\mu$m}
\newcommand{\rsun}{R_{\odot}}
\newcommand{\msun}{M_{\odot}}

\shorttitle{GCIRS 16SW: A Massive Eclipsing Binary}
\shortauthors{Peeples et al.}

\title{The Nature of the Variable Galactic Center Source GCIRS~16SW
Revisited:\\ A Massive Eclipsing Binary}

\author{Molly~S.~Peeples\altaffilmark{1}, A.~Z.~Bonanos\altaffilmark{2},
D.~L.~DePoy\altaffilmark{1}, K.~Z.~Stanek\altaffilmark{1},
J.~Pepper\altaffilmark{1}, Richard~W.~Pogge\altaffilmark{1},
M.~H.~Pinsonneault\altaffilmark{1}, K.~Sellgren\altaffilmark{1} }

\altaffiltext{1}{Department of Astronomy, Ohio State University, 140 W.\
18th Ave., Columbus,~OH~43210} \altaffiltext{2}{Department of
Terrestrial Magnetism, Carnegie Institution of Washington, 5241 Broad
Branch Road, NW, Washington, DC 20015}
\email{molly@astronomy.ohio-state.edu}

\begin{document}

\begin{abstract}
We present a re-analysis of our $H$- and $K$-band photometry and
light-curves for GCIRS~16SW, a regular periodic source near the Galactic
center.  These data include those presented by \citet{depoy04}; we
correct a sign error in their reduction, finding GCIRS~16SW to be an
eclipsing binary with no color variations.  We find the system to be an
equal mass overcontact binary (both stars overfilling their Roche lobes)
in a circular orbit with a period $P=19.4513$~days, an inclination angle
$i=71^{\circ}$.  This confirms and strengthens the findings of
\citet{martins06} that GCIRS~16SW is an eclipsing binary composed of two
$\sim 50\msun$ stars, further supporting evidence of recent star
formation very close to the Galactic center.  Finally, the calculated
luminosity of each component is close to the Eddington luminosity,
implying that the temperature of 24400~K given by \citet{najarro97}
might be overestimated for these evolved stars.
\end{abstract}


\section{Introduction}\label{sec:intro}

Standard star formation modes are thought to break down near a
supermassive black hole (SMBH), raising the question of whether or not
star formation near a SMBH is possible, and if so, through what
mechanism \citep{nayakshin05}.  Our own Galaxy provides us with a unique
opportunity to study individual stars in the presence of a SMBH, namely,
Sgr~A*.  Direct observations of massive, and therefore young, stars
close to Sgr~A* indicate that there has been recent star formation
at the Galactic center \citep{lebofsky82}.

GCIRS~16SW (hereafter IRS16SW) is a variable source near the Galactic
center ($\alpha = 17^{\rm h}45^{\rm m}40\fs 1$, $\delta = -29^{\circ} 00'
29''$, J2000.0). \citet{ott99} reported that the source is regularly
variable and suggested that it could be a binary star with very massive
components.  \citet{depoy04} confirmed the period of the object, but
argued that the source was more likely a pulsating variable. Recently,
however, \citet{martins06} reported spectroscopic observations of
IRS16SW that showed radial velocity variations consistent with a
binary composed of two massive stars.

Prompted by the convincing nature of the radial velocity variations seen
by \citet{martins06} we have re-analyzed the data presented by
\citet{depoy04} as well as additional data from the same observing
campaign. We find that the original data reduction process was seriously
flawed. In particular, the color variation, light-curve asymmetry, and
sign of the brightness variations that \citeauthor{depoy04}\ presented
are artifacts of the data reduction process.

In this Letter, we report on the re-analysis of the
\citeauthor{depoy04}\ data. We find that there is no color change in
IRS16SW over its variations and that the shape of the light curve is
consistent with an eclipsing binary system. The new results are
consistent with \citet{martins06} and confirm that IRS16SW is a binary
composed of two massive stars.  In \S\ref{sec:obs} we describe the
observations and present the data, in \S\ref{sec:analysis} we describe
the best-fit model to the light-curve, and in \S\ref{sec:conc} we
discuss and summarize the results.

\section{Observations}\label{sec:obs}
Observations of the Galactic center in the $H$ (1.6\um) and $K$ (2.2\um)
bands were made at the Cerro-Tololo Inter-American Observatory
(CTIO)/Yale 1-m telescope using the facility optical/infrared imager
(ANDICAM; see \citealt{depoy03} for details).  ANDICAM has a pixel scale
of $0\farcs 22$~pix$^{-1}$ on a $1024\times 1024$ array.  Both $H$- and
$K$-band images were taken in the 2001 and 2002 observing seasons;
$H$-band data were also obtained in 2000. (The \citealt{depoy04} analysis
includes only the 2001 data.)  The observing campaign consists of every
usable night from UTC 2000 August 13 (HJD 2451769.5) through UTC 2000
October 14 (HJD 2451831.5), UTC 2001 May 20 (HJD 2452049.5) through
UTC 2001 November 3 (HJD 2452216.5), and UTC 2002 June 9 (HJD
2452434.5) through UTC 2002 September 25 (HJD 2452542.5).  Each night, a
set of seven slightly offset images were obtained and then combined and
trimmed to form a final nightly image.  The $H$-band images consist of
30~s exposures, and it took about four minutes to obtain the group of
seven images; the $K$-band images consist of 10~s exposures and took
about two minutes to obtain.

The final $512 \times 512$ pixel images, corresponding to a field of
view of $112 \times 112$ arcseconds, are approximately centered on the
Galactic center.  After image quality cuts were made, there are a total
of 144 $H$-band and 137 $K$-band images.  The seeing ranges from
$0\farcs 93$ to $1\farcs 93$ full-width half maximum; in general, the
$H$-band images are of higher quality (typical seeing $\sim 1\farcs 3$)
than the $K$-band (with typical seeing $\sim 1\farcs 45$).

Because the field is crowded, we reduced the data using the ISIS
difference image analysis package \citep{alard00,hartman04}.  This
analysis revealed a sign error in the original \citet{depoy04}
reduction; IRS16SW is clearly an eclipsing binary.  Because IRS16SW is
subject to significant blending---there are roughly half a dozen sources
in the \citet{ott99} catalog within about one arcsecond of IRS16SW---we
calibrated our light-curves with the \citet{ott99} data as presented by
\citet{martins06}.  The \citet{martins06} $K$-band photometry gives a
mean magnitude of 0.2~mag higher than the \citet{ott99} mean, and
includes two more seasons of data.

\begin{figure}
\plotone{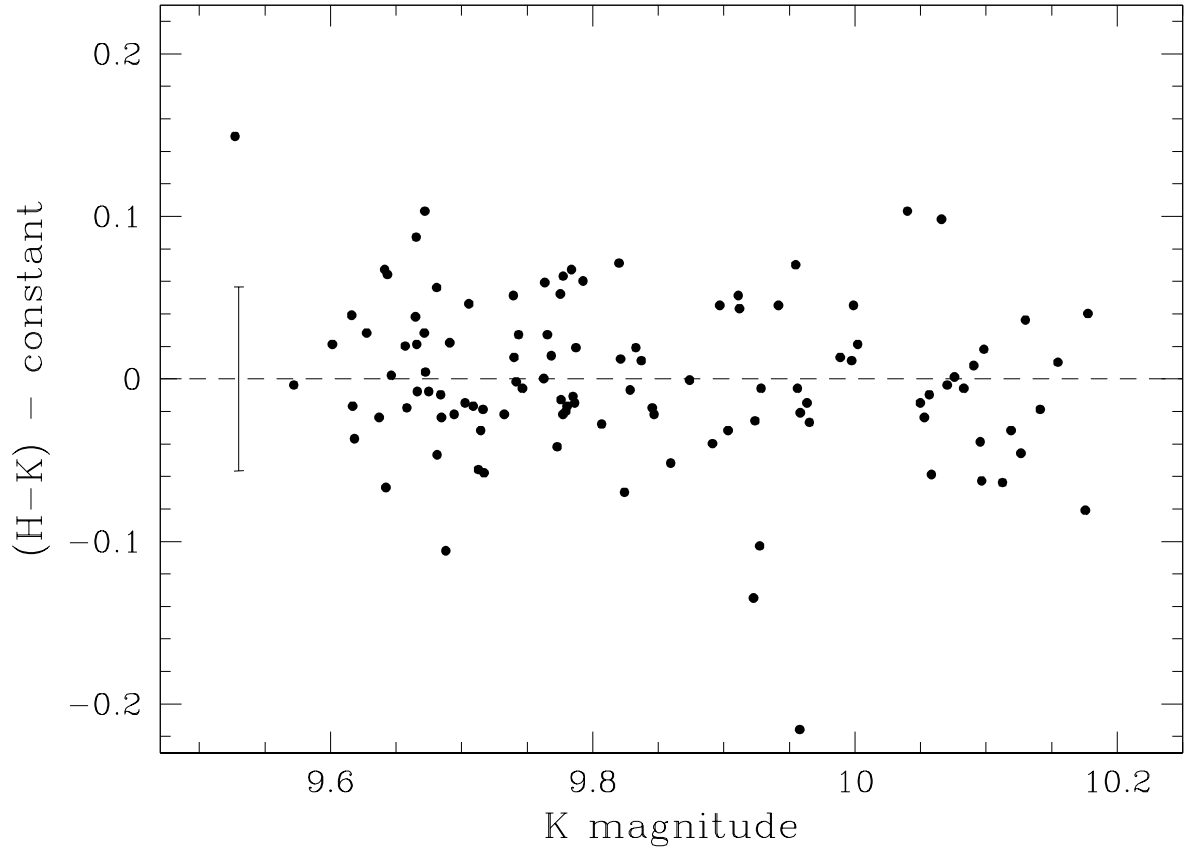}
\caption{\label{fig:color} $H-K$ color residuals for DAOPhot photometry
  versus calibrated $K$-band magnitude. No clear trend between color and
  magnitude is observed.  The rms variation about a constant color is
  0.05~magnitudes.  The errorbar on the left shows the typical
  uncertainty in $H-K$ color for a constant star of similar magnitude.
  See \S~\ref{sec:obs} for further discussion.}
\end{figure}

\citeauthor{ott99}\ found a variability amplitude of 0.55~mag; using
DAOPhot photometry to scale the ISIS fluxes, we find an amplitude of
$\sim 0.35$~mag in both $H$ and $K$ \citep[see][Appendix B]{hartman04}.
This substantial amplitude difference is indicative of significant
blending in our data.  We used the period of 19.45~days reported by
\citet{depoy04} and \citet{martins06} to scale our $K$-band light-curve
to have the same mean magnitude and amplitude as the
\citeauthor{martins06}\ data.  There were 110 nights for which data were
obtained in both the $H$- and $K$-bands, providing contemporaneous
measurements of the $H-K$ color.  Using the DAOPhot photometry, we find
a constant $H-K$ color with an rms of 0.05~mag; this color does not vary
with time, phase, $H$-, or $K$-band magnitude, as shown in
Figure~\ref{fig:color}.  Lacking properly calibrated $H$-band data, we
scaled the similarly blended $H$-band data to have the same amplitude
and mean magnitude as the $K$-band data.  These scaled light-curves form
the basis of our analysis; Table~\ref{tbl:data} gives the final scaled
time-series photometry.

\section{Light Curve Analysis}\label{sec:analysis}
We simultaneously fit the 144 $H$-band and 184 $K$-band points
(including the \citeauthor{martins06}\ $K$-band points, except for two
noisy points at HJD 2498704 and 2499908) using the October 2005 version
of the Wilson-Devinney (WD) code \citep{Wilson71, Wilson79, wilson90} in
the overcontact mode (MODE 3). We fixed $T_{\rm eff1}=24400$~K as
estimated by \citet{najarro97} and used the square-root limb darkening
law, taking the values of the limb darkening coefficients from \citet
{claret00} for a LTE ATLAS9 \citep{kurucz93} stellar atmosphere model
with $T_{\rm eff}=24000$~K, $\log(g)=3.0$ (cgs), turbulent velocity of
2~$\rm km \; s^{-1}$ and solar metallicity. We fixed gravity brightening
exponents and albedos to unity from theoretical values for stars at such
temperatures.  We assumed equal masses, circular orbits, and synchronous
rotation, fitting for 7 parameters: the period $P$, time of primary
eclipse $T_{\rm prim}(\rm HJD)$, inclination $i$, $T_{\rm eff2}$, the
luminosity of the primary in each band ($L_{1H},L_{1K}$) and the surface
potential ($\Omega_{1}=\Omega_{2}$, \citealp[see][eq.~1]{Wilson79}).
We defined convergence to be when the corrections for all adjusted
parameters were smaller than their respective standard or statistical
errors after three consecutive iterations. The best fit parameters are
shown in Table~\ref{tbl:WD}. The ephemeris is
\begin{equation}
T_{\rm prim}=2451775.102 \pm 0.032 + 19.4513 \pm 0.0011 \times E\; (\rm HJD).
\end{equation}

A good fit required that the stars overfill their limiting Roche lobes,
which justifies using the overcontact mode of WD. \citet{martins06}
adopted the largest filling factor allowed by NIGHTFALL (1.3); we
calculate a larger fill-out factor (as defined by \citealt{mochnacki72})
of $F=1.44$. The critical surface potentials for the inner and outer
surfaces under the above assumptions are $\Omega_{\rm in}=3.75$ and
$\Omega_{\rm out}=3.21$. Assuming different values for the mass ratio
$q$ also produced good fits. However, a photometric mass ratio is not
well constrained by our photometry; therefore, we did not attempt to
solve for it.  The inclination we derive is $i=71$\dg, in agreement with
\citet{martins06}; however, if there is still unaccounted for blending,
then the inclination angle could be larger.

\begin{figure}
\plotone{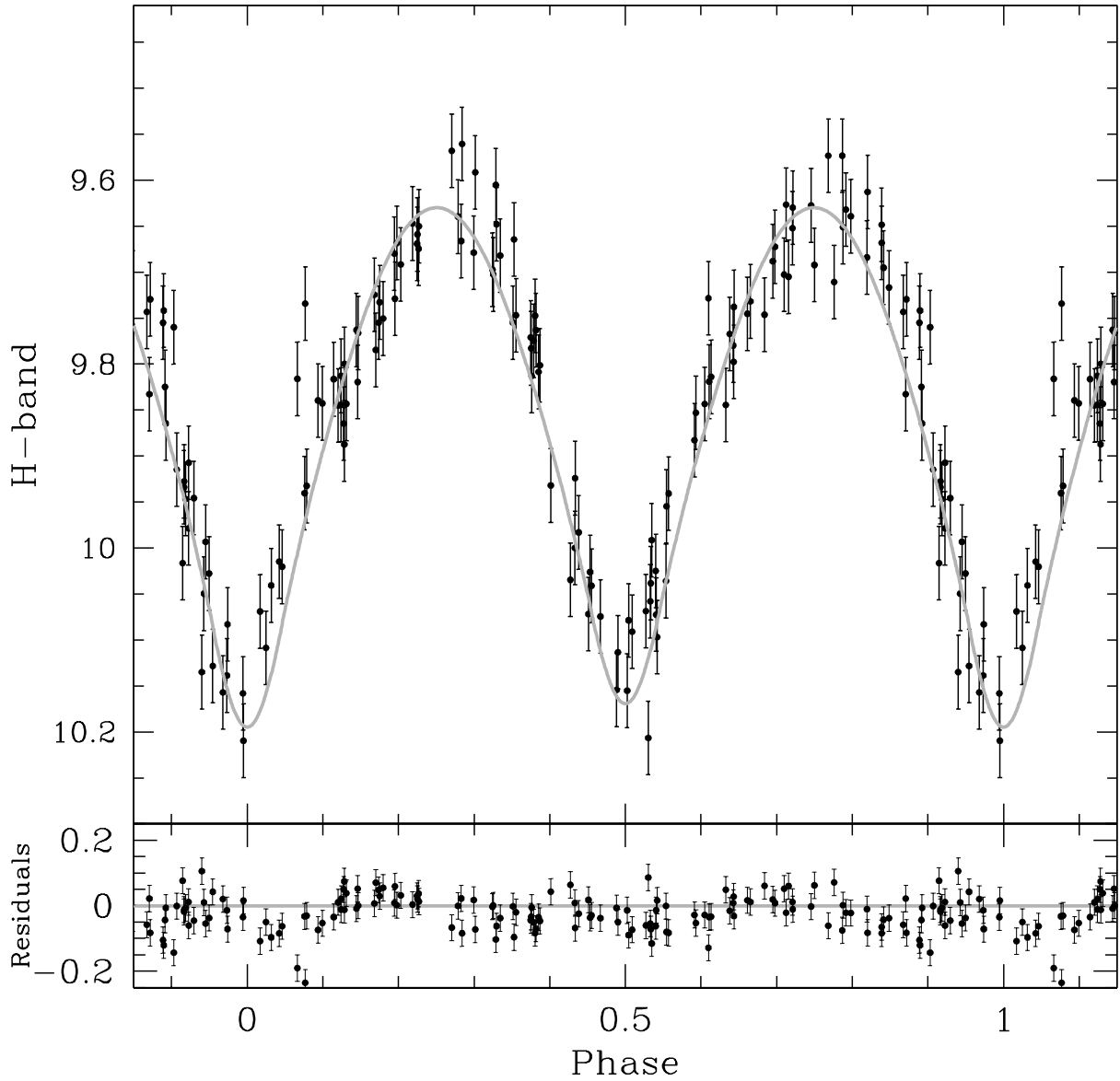}
\caption{\label{fig:lch}Wilson-Devinney fit and residuals of an
  overcontact binary to the $H$-band light-curve of IRS16SW.  The period
  is 19.4513~days; the model parameters have zero eccentricity and an
  inclination of 71\dg.  The rms variation about the model fit is
  0.06~magnitudes. }
\end{figure}

\begin{figure}
\plotone{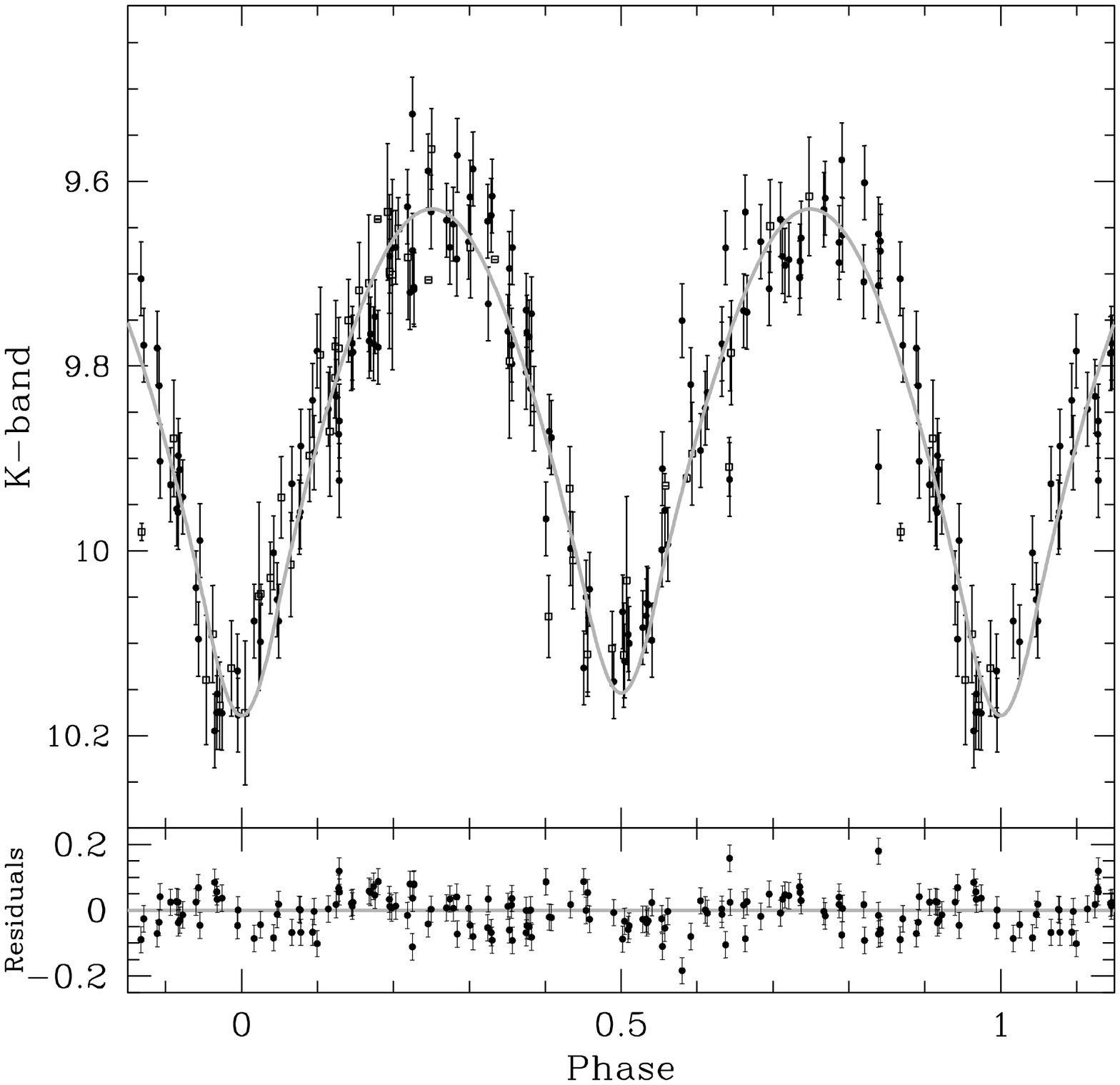}
\caption{\label{fig:lck}Wilson-Devinney fit and residuals of an
  overcontact binary to the $K$-band light-curve of IRS16SW.  The filled
  circles are the data presented here; the open squares are the
  \citet{martins06} data.  The period is 19.4513~days; the model
  parameters have zero eccentricity and an inclination of 71\dg. The rms
  variation about the model fit is 0.06~magnitudes.}
\end{figure}

In Figures~\ref{fig:lch} and \ref{fig:lck} we show the $H$- and $K$-band
light curve model fit for IRS16SW. The \citet{martins06} data are
plotted with unfilled symbols in Figure~\ref{fig:lck}. Error bars for
our data are set to 0.04~mag, corresponding to the typical variation
seen for a constant star of similar magnitude (Peeples et al., in
preparation).  The fact that the data are fit well under the assumption
of circular orbits is an indication that this assumption is sound; these
data give no evidence for an eccentric orbit.  The eccentricity of
$e=0.09$ found by \citet{martins06} is derived from their radial
velocity curve, which tends to yield nonzero eccentricities
\citep{lucy71}.  Furthermore, the circularization time for a system with
these physical characteristices (discussed below) is only tens of
thousands of years \citep{zahn75, zahn77}, making it unlikely that we
are observing IRS16SW pre-circularization.

\begin{figure}
\plotone{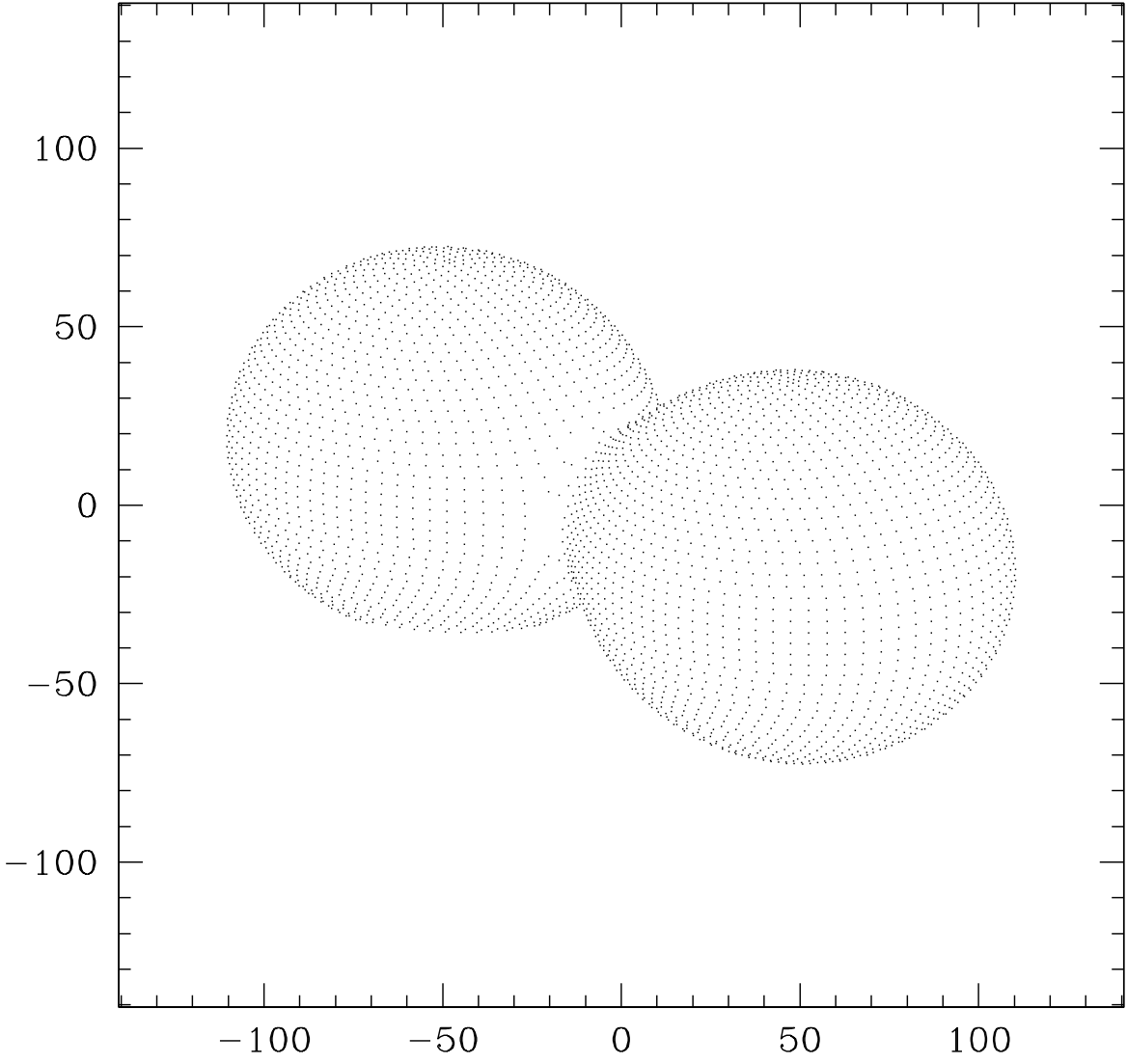}
\caption{\label{fig:peanut}Wilson-Devinney visualization of IRS16SW
  at an orbital phase of 0.12. Axes are in units of $\rsun$.}
\end{figure}

From the definition of $\Omega$, WD calculates the following best fit
fractional radii for both stars in units of the orbital separation: the
polar radius, $r_{\rm pole}=0.39$; the radius in the plane of revolution
and perpendicular to the line connecting the stars' centers, $r_{\rm
side}=0.41$, and the radius in the direction of L2, $r_{\rm
back}=0.47$. The orbital separation of \citet{martins06}, $(a_1+a_2)\sin
i = 132.8 \pm 4.4\,\rsun$, yields physical radii of $R_{\rm pole}=54.5
\pm 1.8\, \rsun$, $R_{\rm side}=58.2 \pm 1.9\, \rsun$, and $R_{\rm
back}=62.7 \pm 2.1\,\rsun$.  The WD visualization for this system is
shown in Figure~\ref{fig:peanut}.

\section{Discussion and Conclusion}\label{sec:conc}

Using Kepler's law, \citet{martins06} find $M_1 \approx M_2 \approx
50\msun$, placing the components of IRS16SW among the most massive stars
known.  Until recently, the most massive stars measured in binaries were
R136-38 in the Large Magellanic Cloud \citep[$57\;\msun$,][]{massey02}
and WR~22 ($55\;\msun$, \citealt{rauw96}, \citealt{schweickhardt99}), an
evolved star in our Galaxy. The current heavyweight champion is a
Wolf-Rayet binary, WR~20a (82 \& 83 $\msun$, \citealt{rauw04},
\citealt{bonanos04}), in the young Galactic cluster Westerlund~2.

The luminosity of IRS16SW poses a problem.  Using a radius of
$R=59.7\rsun$ (the mean radius given by WD for an orbital separation of
$140.6\rsun$) and an effective temperature of $T_{\rm eff}=24400$~K
\citep{najarro97}, we can estimate the luminosity $L=4\pi R^2 \sigma
T_{\rm eff}^4$ of each component as $4.4\times 10^{39}\,\rm erg \; s
^{-1}$.  The non-sphericity of IRS16SW will only drive this luminosity
higher.  For comparison, the Eddington luminosity of a $50\msun$ star is
$L_{\rm edd} = 1.3 \times 10^{38} (M/\msun) = 6.5 \times 10^{39}\,\rm
erg \; s ^{-1}$.  It is highly unlikely that each component of IRS16SW
has been radiating stably at nearly their Eddington luminosities for
eleven years \citep{humphreys94}; the combined photometry of
\citet{ott99} and this work span 1992--2002.  Assuming the orbital
separation as calculated by \citet{martins06} is correct (if it is
smaller, then $L/L_{\rm edd}$ will be even larger), this calculation
implies that the temperature of \citet{najarro97} is an over-estimate.
A change in the assumed $T_{\rm eff}$ affects the WD model parameters;
specifically, a decrease in $T_{\rm eff}$ by a few thousand degrees
Kelvin will decrease the inclination angle $i$, and thus increase the
masses of the stars, by more than the formal $1\sigma$ uncertainties
given by WD.

A radial velocity curve for the secondary is necessary to determine the
value of the mass ratio $q$.  It remains a puzzle as to why IRS16SW does
not appear to be a double-line spectroscopic binary.  It is readily
apparent from the depths of the eclipses that the two stars have
near-equal fluxes and from the depth ratio that they have near-equal
surface brightnesses, yet \citet{martins06} see only one set of
spectroscopic lines.  However, since the spectral features used by
\citet{martins06} are wind lines, with strong characteristic P~Cygni
profiles \citep{najarro97}, differences in wind strength or small
differences in the effective temperatures of the stars could easily
conspire to make detection of the second set of lines difficult.

We confirm that GCIRS~16SW is a massive eclipsing binary with both stars
overflowing their Roche lobes.  We find a refined orbital period of
$19.4513 \pm 0.0011$~days and an inclination of 71\dg\ with an assumed
mass ratio of 1, supporting the findings by \citet{martins06} that the
masses of the two stars are both $\sim 50\msun$.  The projected distance
between IRS16SW and Sgr~A* is 0.05~pc $\sim 11000$~AU (assuming a
distance to the Galactic center of 7.6~kpc, \citealt{eisenhauer05}); in
fact, IRS16SW is part of a moving group that is likely bound to Sgr~A*
\citep{lu05, paumard06}.  As the lifetime of $50\msun$ stars is $\sim
4$~Myr \citep{schaller92}, these observations are strong evidence that
IRS16SW was formed within 0.1~pc of Sgr~A* despite the tidal shear from
the black hole which creates problems in star formation models.

\acknowledgements 
We thank F.\ Martins for providing us with the \citet{ott99} $K$-band
light-curve, Slavek Rucinski, Andy Gould, and John Beacom for useful
discussions, and the anonymous referee for helpful suggestions.
A.~Z.~B. acknowledges research and travel support from the Carnegie
Institution of Washington through a Vera Rubin Fellowship. KS gratefully
acknowledges support from NSF grant AST-0206331.

\clearpage

\begin{deluxetable}{crr}
\tablewidth{0pc}
\tablecaption{\sc $H$- and $K$-band Photometry of GCIRS~16SW
\label{tbl:data}
}
\tablehead{
\colhead{Band} & 
\colhead{HJD} &
\colhead{Scaled} \\
\colhead{} &
\colhead{$-2450000.$} &
\colhead{Magnitude}}
\startdata
 H  & 1769.6620  & 9.630 \\  
 H  & 1772.5868  & 9.730 \\  
\dotfill & \dotfill & \dotfill \\
 H  & 2542.5417  & 10.041\\  \hline
 K  & 2048.9194  & 9.886 \\  
 K  & 2051.7765  & 9.527 \\  
\dotfill & \dotfill & \dotfill \\
\enddata

\tablecomments{Both $H$- and $K$-bands are scaled to have the same
  amplitude and mean magnitude as the \citet{ott99} $K$-band data
  presented by \citet{martins06}.  All errors are set to 0.04~mag,
  corresponding to the typical variation seen for a constant star of
  similar magnitude (Peeples et al., in preparation).  Table
  \ref{tbl:data} is published in its entirety in the electronic edition
  of the Astrophysical Journal. A portion is shown here for guidance
  regarding its form and content.  }

\end{deluxetable}

\clearpage

\begin{deluxetable}{lc}
\tablewidth{0pc}
\tablecaption{\sc Best-Fit Parameters From Combined $H$\&$K$ Light-Curve
  Analysis With Wilson-Devinney Program\label{tbl:WD}
}
\tablehead{
\colhead{Parameter} & \colhead{Value}}
\startdata
Period, P & 1$9.4513 \pm 0.0011$~days \\
Time of primary eclipse, $T_{\rm prim}$ & $2451775.102 \pm 0.032$ \\
Inclination, $i$       & $70.85$\dg  $\pm 0.6$\dg\\
Temperature ratio, $T_2 / T_1$ & $0.96$ \\
Surface potential, $\Omega$ & 3.51 \\
Light ratio in $H$, $L_{2}/L_{1}$ & 0.936\\
Light ratio in $K$, $L_{2}/L_{1}$ & 0.939\\
Radius, $r_{\rm pole,\, 1} = r_{\rm pole,\, 2}$& 0.39 \\
............ $r_{\rm side,\, 1} = r_{\rm side,\, 2}$ & 0.41 \\
............ $r_{\rm back,\, 1} = r_{\rm back,\, 2}$ & 0.47 \\
& \\
Secondary temperature, $T_2$ & 23500~K\\
Radius, $R_{\rm pole}$ & $54.5 \pm 1.8\, \rsun$ \\
............ $R_{\rm side}$ & $58.2 \pm 1.9\,\rsun$ \\
............ $R_{\rm back}$ & $62.7 \pm 2.1\,\rsun$ 
\enddata

\tablecomments{First ten parameters are best-fit parameters from a
  combined $H$\&$K$ light-curve analysis with the Wilson-Devinney (WD)
  program.  The $1\sigma$ uncertainties given by WD are unrealistically
  small, and thus not listed.  The radii $r_{\rm pole}$, $r_{\rm side}$,
  and $r_{\rm back}$ are in units of the orbital separation.
  The final four (physical) parameters are
  based on the orbital separation, $(a_1 +a_2)\sin i = 140.6 \pm
  4.7\rsun$ and the assumed effective temperature of 24400~K for $T_1$
  \citep{najarro97}. }

\end{deluxetable}

\clearpage

\clearpage

\clearpage

\clearpage

\end{document}